# Proton scattering power of some tissue-equivalent plastics

V.N. Vasiliev[*,1], V.I. Kostjuchenko[2], V.G. Khaybullin[2]
S.I. Samarin[3], A.S. Uglov[3]

[1]Institute for Nuclear Research, Russian Academy of Sciences, Moscow, Russia
[2]Institute for Theoretical and Experimental Physics, Moscow. Russia
[3]Russian Federal Nuclear Center–Zababakhin Institute of Applied Physics, Snezhinsk, Russia

Proton scattering in some water and tissue equivalent phantom materials was measured to evaluate their simulation accuracy of water and respective human biological tissues. The measurements were performed on the medical facility of the ITEP synchrotron, proton energy was 219 MeV, a narrow beam was formed by a 3 mm collimator. A stack of plastic slabs was set closely to the collimator hole as a scatterer. Three types of Plastic Water (PW, PW LR and PW DT), lung, cortical bone, adipose and muscle plastics (CIRS Inc., USA) were used in the experiments as the substitutes under investigation and liquid water and PMMA slabs as reference materials. Dose (intensity) profiles were measured for each sample by two orthogonal strips of the Gafchromic EBT film. A total thickness of the plastic slab was from 4 to 16 cm depending on the material. The Gafchromic film response nonlinearity was taken into account by an additional calibration vs. absorbed dose in a wide proton beam, the temporal irradiation-to-scanning dependence was also accounted. The central part of each angular distribution was fitted by the Gaussian function and compared with the respective parameters calculated for the simulated medium by Monte Carlo technique with the IThMC code.

## Introduction

In our previous paper [1], a number of water and tissue equivalent materials originally developed as substitutes for radiation therapy with photon and electron beams were investigated to evaluate their applicability in proton therapy. The proton CSDA ranges were measured and compared with those of respective human tissues and water, thus, stopping power ratios of these materials also were evaluated indirectly. In this work, we have measured another significant parameter, scattering power, affecting on proton transport in phantoms and test objects. The obtained results were compared with theoretical estimates simulated by Monte Carlo technique.

## Materials and methods

The measurements of proton scattering in the materials under investigation were performed on the ITEP synchrotron medical beam facility (Fig.1). Initial energy of proton beam was 219 MeV, a double scattering system for medical beam spreading as well as a water energy degrader were removed from the experimental arrangement. A narrow proton beam was formed by a steel collimator of diameter 3.0 mm.

A scattered protons detection plane was at a distance of 802 mm from the collimator. Two orthogonal strips of Gafchromic EBT radiochromic film, lot #36347-02I, was placed at the plane and formed a cross at the beam axis. Along with laser pointers, the beam axis alignment was additionally verified by a direct measurement with a piece of the film, without a scatterer.

The samples investigated in the measurements are listed in Table 1. Total thickness of each material has varied from 40 to 160 mm, the slabs were set close to the collimator as shown in Fig. 1. Along with the plastics, a PMMA slab and liquid water in a cuvette were included in the measurements as reference media for an independent estimation of the Monte Carlo simulation accuracy.

---

[*] E-mail: vnvasil@orc.ru

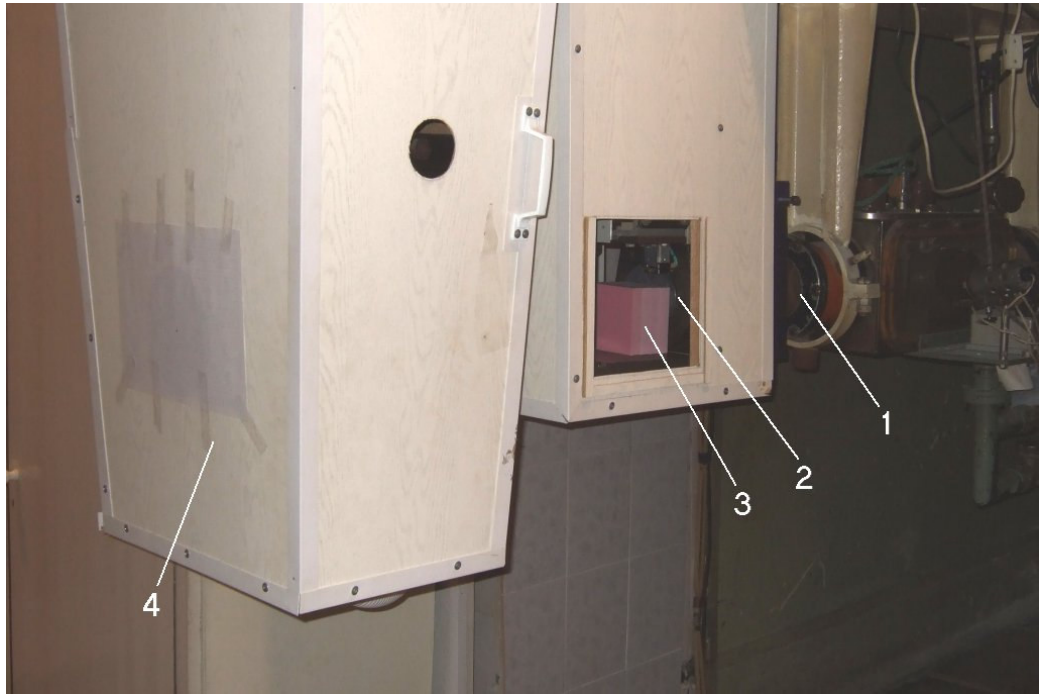

Fig.1. The experimental setup. 1 – a proton vacuum transport system, 2 – a steel collimator; 3 – a material under investigation; 4 – a detection plane.

Table 1. The substitutes under investigation.

| Material | Density, g/cm$^3$ | Thickness, mm |
|---|---|---|
| Plastic Water | 1.030 | 80 |
| Plastic Water LR | 1.029 | 80 |
| Plastic Water DT | 1.039 | 80 |
| Lung | 0.205 | 160 |
| Cortical Bone | 1.91 | 40 |
| Adipose | 0.96 | 80 |
| Muscle | 1.04 | 80 |

The exposed film strips were scanned by a flatbed scanner Epson 4990 Photo in the transparence mode. The scanner ensured a 4D optical density interval, all images were obtained and stored as 48 bit/pixel TIFF files without compression. A piece of unexposured film was scanned at the same image to calculate optical density difference by following equation

$$OD = \log_{10}\left(\frac{I_0}{I_D}\right),$$

where $I_D$ is a pixel value for an exposed piece of film; $I_0$ is the same for unexposured film. The calibration curve "absorbed dose – optical density" was obtained earlier using the same lot of the Gafchromic EBT film. The optical density profiles were then converted to radial absorbed dose profiles and fitted by a Gaussian function. The radial parameter $\sigma_r$ of the Gaussian was used for results comparison in further analysis.

Along with the experiment, angular and radial distributions of scattered protons were simulated by Monte Carlo technique. The proton transport simulation was performed by the program IThMC developed for proton therapy planning and allowing a dose calculation in voxel geometry (up to 512x512x512 voxels). The program takes into account ionization energy loss in the medium, energy straggling (by the Landau, Vavilov or normal distributions depending on the material thickness), elastic multiple coulomb scattering using the Fokker-Planck and Fermi-Eyges models, elastic and inelastic nuclear reactions on the base of the Sychev model and the D2N2 cross section data set respectively [2]. In these calculations, a Gaussian model of multiple proton scattering was used, at least $5 \times 10^8$ incident protons were sampled for each material under investigation.

## Results and discussion

### Reference materials: experiment vs. Monte Carlo

In our measurements, water and PMMA were used as reference materials with well established properties to verify the simulation of multiple proton scattering in the Monte Carlo code. The respective profiles are presented in Fig. 2 and the fitted parameter $\sigma_r$ is shown in Table 2. In addition, the results without any sample are presented to estimate scatter contributions in the collimator and air.

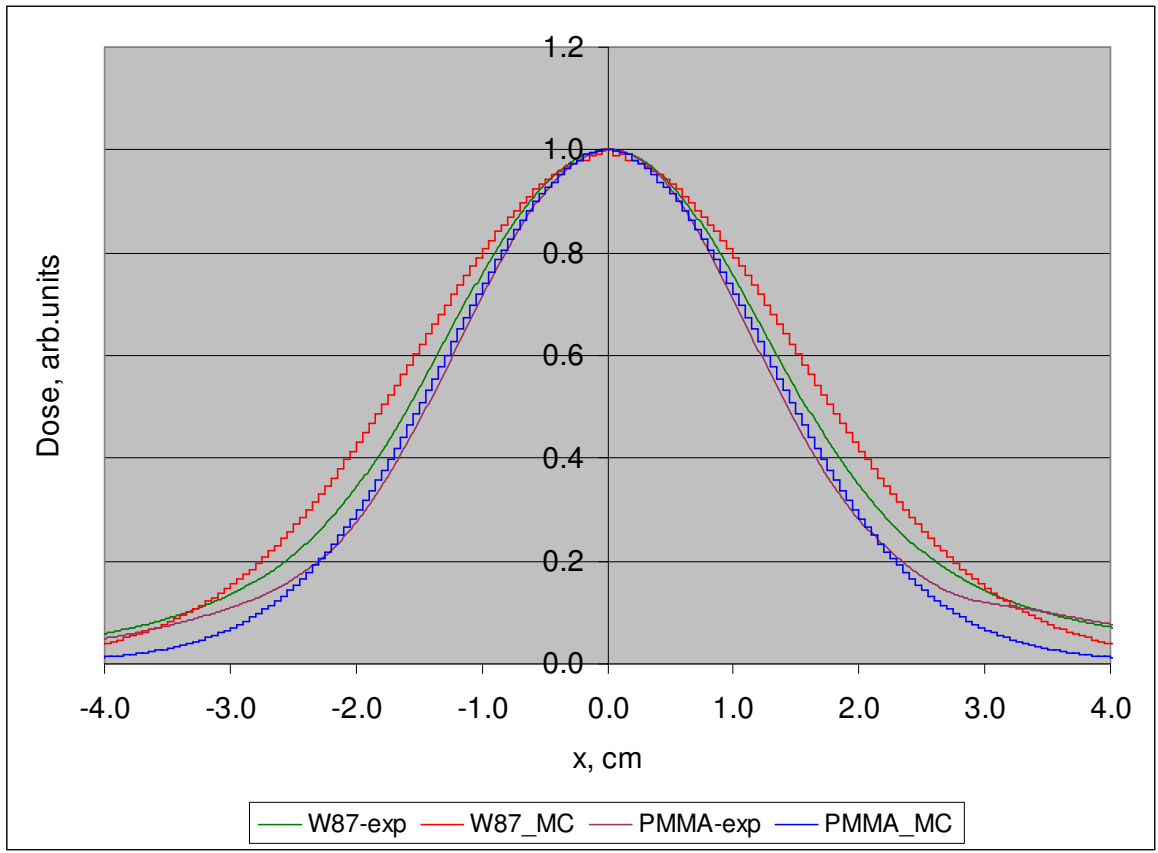

Fig. 2. Experimental and simulated dose profiles after scattering in water, 8.75 cm, and PMMA, 5.82 cm. The histograms are Monte Carlo data.

Table 2. A fitted Gaussian parameter $\sigma_r$ for the reference media and the collimator hole.

| Material | EXP or MC | Profile direction | Thickness, cm | $\sigma_r$, cm |
|---|---|---|---|---|
| No | EXP | Horizontal | - | 0.13 |
| No | EXP | Vertical | - | 0.09 |
| No | MC | - | - | 0.13 |
| Water | EXP | Horizontal | 8.75 | 1.34 |
| Water | MC | - | 8.75 | 1.53 |
| PMMA | EXP | Horizontal | 5.82 | 1.19 |
| PMMA | MC | - | 5.82 | 1.27 |

Measured in horizontal direction and simulated profiles obtained for open collimator hole, without a sample, are in excellent agreement; the resulting FWHM is 3.0 mm and equal to the collimator diameter. This relation demonstrates little contribution of scatter in air between the collimator and the detection plane.

The experimental profile of the collimator hole in vertical direction is slightly less than in horizontal one. That fact can be resulted from a little misalignment of the collimator axis, insignificant in medical use of the proton beam but detectable with the collimator diameter decrease. Other experimental results also demonstrate that the vertical profile is a little more narrow (within 1 mm). To eliminate this influence, namely horizontal profiles are referred in our further analysis as the experimental data.

A comparison of the water and PMMA results (Table 2) demonstrates the measured data are systematically lower by 0.8-1.9 mm (7-14%). This difference is statistically significant and, perhaps, resulted from a little misadjustment of scattering parameters in the Monte Carlo transport model. It should be accounted for further results comparisons.

**Substitutes vs. tissues comparison: Monte Carlo results**

A scatter in three types of water equivalent substitute, Plastic Water, Plastic Water LR and Plastic Water DT, as well CIRS's lung inhale and cortical bone were simulated by Monte Carlo technique in geometry identical to used in the experiment. Similar calculations were performed with reference data, ICRU compact bone, ICRP cortical bone and W-W lung, using ICRU 49 [3] and Woodard and White [4] elemental compositions. Resulting dose profiles had exact Gaussian shape according to the used model of multiple proton scattering; their parameters $\sigma_r$ are presented in Table 3.

Table 3. Gaussian parameters $\sigma_r$ derived from Monte Carlo calculated dose profiles.

| Material | Thickness, cm | $\rho$, g/cm$^3$ | $\sigma_r$, cm | Substitute/reference ratio |
|---|---|---|---|---|
| **PW** | 8.0 | 1.03 | 1.46 | 1.000 |
| **PWLR** | 8.0 | 1.029 | 1.43 | 0.982 |
| **PWDT** | 8.0 | 1.039 | 1.45 | 0.993 |
| Water | 8.0 | 1.000 | 1.46 | |
| **CIRS Cortical bone** | 4.0 | 1.91 | 1.67 | 1.071 (ICRU), 1.018(ICRP) |
| ICRU Compact bone | 4.0 | 1.85 | 1.56 | |
| ICRP Cortical bone | 4.0 | 1.85 | 1.64 | |
| **CIRS Lung inhale** | 16.0 | 0.205 | 0.78 | 0.963 |
| WW Lung | 16.0 | 0.200 | 0.81 | |

Very good agreement was demonstrated between water and its substitutes. Scattering power of Plastic Water is identical to that of liquid water, the difference of PWDT and PWLR scatter never exceeded 1.8%. Similar little disagreement was found for cortical bone substitute in comparison with ICRP cortical bone. Nevertheless, bone elemental composition and physical density differ depending on the reference data source. In particularly, the difference with ICRU compact bone amounts to 7%. The lung substitute shows a 3.7% underestimation of scattering power against the respective tissue.

**Substitutes comparison: experiment vs. Monte Carlo**

In this section, a comparison is presented of the scatter parameters measured with the substitutes and calculated in liquid water and the respective tissues. The obtained results are shown in Fig. 3, 4, and Table 4. The water substitutes demonstrate very good data agreement within 1.8%. The results of all tissue substitutes are from 0.4 to 1.2 mm lower than Monte Carlo results for respective biological tissues. Thus, we have approximately equal discrepancy as was observed in water and PMMA comparison and was related to the imperfection of the Monte Carlo model as discussed above.

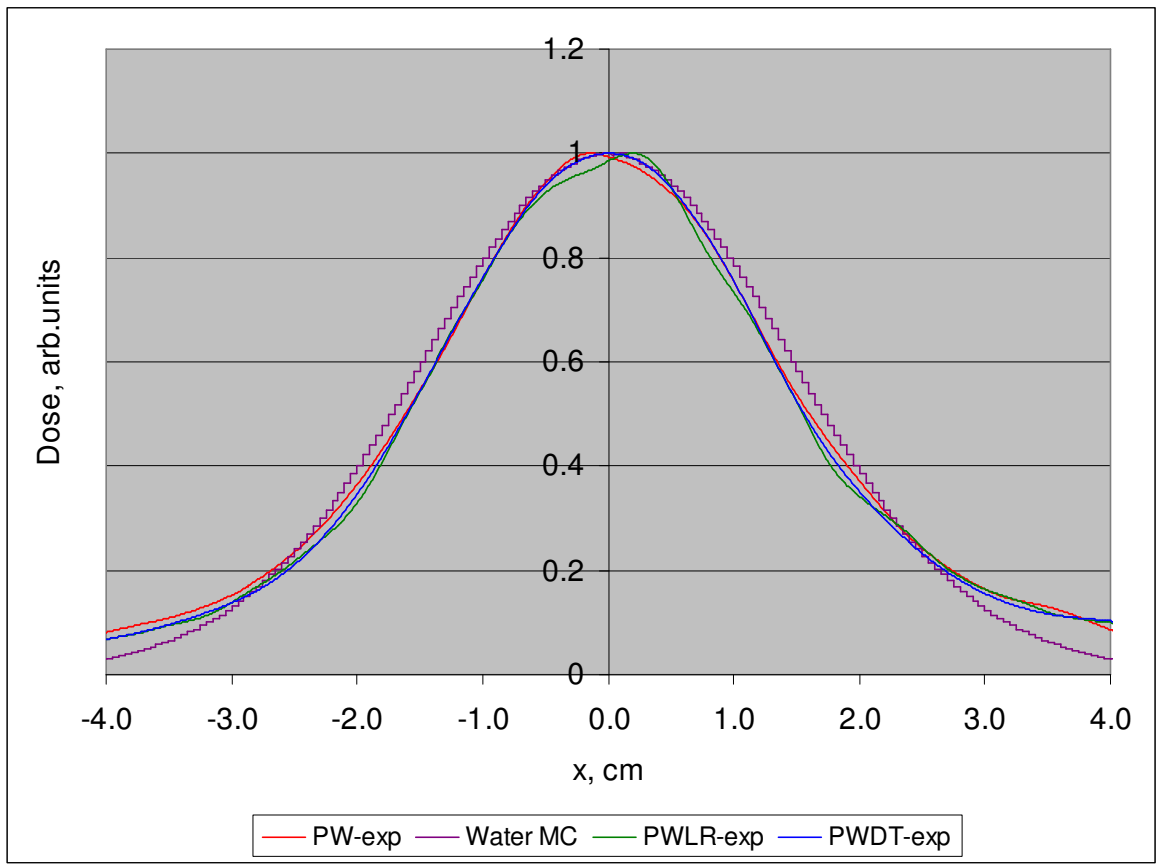

Fig. 3. A comparison of dose profiles measured with water substitutes and calculated with liquid water.

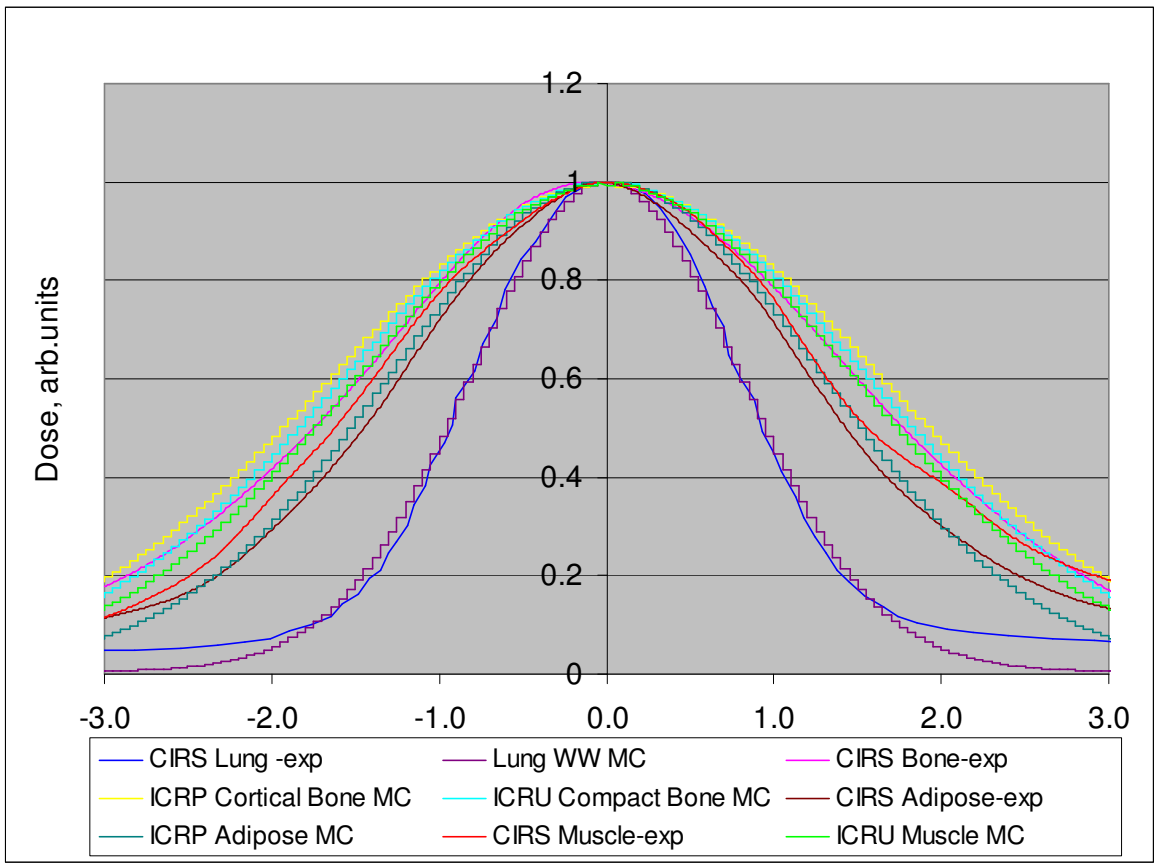

Fig. 4. A comparison of dose profiles measured with tissue substitutes and calculated with respective tissues.

Table 4. A comparison of dose profiles parameters measured with tissue substitutes and calculated with respective tissues.

| Material | EXP or MC | Thickness, cm | $\sigma_r$, cm | Substitute/reference ratio |
|---|---|---|---|---|
| **PW** | EXP | 8.0 | 1.46 | 1.000 |
| **PWLR** | EXP | 8.0 | 1.43 | 0.982 |
| **PWDT** | EXP | 8.0 | 1.45 | 0.993 |
| Water | MC | 8.0 | 1.46 | |
| **CIRS Cortical bone** | EXP | 4.0 | 1.52 | 0.974 (ICRU), 0.927 (ICRP) |
| ICRU Compact bone | MC | 4.0 | 1.56 | |
| ICRP Cortical bone | MC | 4.0 | 1.64 | |
| **CIRS Lung inhale** | EXP | **16.0** | **0.76** | **0.938** |
| WW Lung | MC | 16.0 | 0.81 | |
| **CIRS Adipose** | EXP | **8.0** | **1.22** | **0.946** |
| ICRP Adipose | MC | 8.0 | 1.29 | |
| **CIRS Muscle** | EXP | **8.0** | **1.23** | **0.831** |
| ICRP Skeletal Muscle | MC | 8.0 | 1.48 | |
| ICRU Striated Muscle | MC | 8.0 | 1.48 | |

The only substitute that demonstrated a disagreement as great as 17% (2.5 mm) is muscle. So large difference can not be explained by the features of the MC calculation model and indicates a significant underestimation of scattering power in this plastic, at least, 10-15%.

The obtained imperfection of the muscle substitute, perhaps, can not result in a significant dose error in a wide proton beam, typical for medical applications, where a lateral multiple scatter equilibrium exists. In particularly, this does not restrict a use of the muscle substitute in human phantom manufacturing and proton beam dose measurements. Nevertheless, we can state a few cases where underestimation of scattering power can be detectable:

- measurements of the proton beam penumbra with a high spatial resolution detector (for example, film);
- dose measurements of very narrow proton beams, about 1-2 mm in diameter, where lateral scatter equilibrium is missing and depth dose distribution significantly depends on the lateral scatter leakage;
- scatter measurements in conditions of a long throw (a distance between the scatterer and the detection plane) – like used in our measurements.

Scatter discrepancy of other substitutes correspond to typical uncertainty of scattering power calculation by various theoretical models [5] and are quite adequate to use in human phantoms and test objects.